\documentclass[12pt]{article}
\begin{document}
\baselineskip=20pt

\title{Bulk versus boundary quantum states}

\author{\large Henrique Boschi-Filho\footnote{\noindent e-mail: 
boschi @ if.ufrj.br}\,  
and 
Nelson R. F. Braga\footnote{\noindent e-mail: braga @ if.ufrj.br}
\\ 
\\ 
\it Instituto de F\'\i sica, Universidade Federal do Rio de Janeiro\\
\it Caixa Postal 68528, 21945-970  Rio de Janeiro, RJ, Brazil}
 
\date{}

\maketitle

\vskip 3cm

\begin{abstract}

An explicit holographic correspondence between $AdS$ bulk and boundary quantum states 
is found in the form of a one to one mapping between scalar field
creation/annihilation operators.
The mapping requires the introduction of arbitrary energy scales and
exhibits an ultraviolet-infrared duality: a small regulating 
mass in the boundary theory corresponds to a large momentum cutoff in the bulk.
In the massless (conformal) limit of the boundary theory the mapping covers the whole 
field spectrum of both theories.
The mapping strongly depends on the discretization of the field spectrum of
compactified $AdS$ space in Poincare coordinates. 

\end{abstract}


\vfill\eject


The holographic principle asserts that  a 
quantum system with gravity can be represented by  a theory on the 
corresponding boundary\cite{HOL1,HOL2,HOL3}.
This principle was inspired by the  result that the black hole
entropy is proportional to its horizon area\cite{BHE1,BHE2}.
A realization of that principle was proposed by Maldacena in the form of a 
conjecture\cite{Malda} on the equivalence (or duality) of the large 
$N$ limit of $SU(N)$ superconformal field theories in $n$
dimensions with supergravity (as a limit of superstring theory) defined in 
$n+1$ dimensional anti de Sitter spacetime
times a compact manifold $\,(AdS/CFT\,$ correspondence). 
Prescriptions for realizing this conjecture, using Poincar\'e 
coordinates in the $AdS$ bulk, were established by  
Gubser, Klebanov and Polyakov \cite{GKP} and  Witten \cite{Wi}. 
In their approach, the $AdS$ solutions play the role of 
classical sources for the boundary field correlators (for a review and a wide list
of references see\cite{Pe,Malda2}).
The relation between the holographic mapping and the 
renormalization group flow was discussed in \cite{BK}.
Further, the recent model of Randall and Sundrum\cite{RS} that proposes a solution
to the  hierarchy problem also presents holographic mapping between $AdS$ bulk and 
boundary\cite{Ve}.

The isomorphism between  the Hilbert spaces of the AdS string theory and
the boundary CFT was established in \cite{HS1,HS2,HS3,HS4}.
However, in this context it is difficult to find an explicit one to one mapping
between bulk and boundary quantum states.
Besides the involving string structure, one source of difficulty for an explicit 
mapping is the different dimensionality of the spaces.
So it would be interesting to have an example of a one to one mapping 
between bulk and boundary quantum states. We show that this is possible 
considering a simple model with scalar fields for bulk and boundary.
Scalar fields in the AdS bulk has already been discussed in\cite{GKP,Wi},
although the associated boundary field in the AdS/CFT correspondence would be composite 
as can be seen from its conformal dimension.

In this letter we find an explicit one to one relation  between the 
creation-annihilation operators of scalar fields  in $AdS$ spacetime and on 
its boundary. This implies a direct relation between the corresponding quantum states.
This mapping is possible because of a discretization of the field spectrum in
the AdS bulk as discussed previously in refs.\cite{BB1,BB2}.
A fundamental ingredient is that canonical commutation relations in both theories 
are preserved. This is a realization of the holographic principle.
One remarkable fact is that there is an ultraviolet-infrared duality.
Starting with boundary fields with some small mass $\mu$ (that can be interpreted
as some infrared regulator) we find that the bulk field has an ultraviolet
cut off behaving as $\,1/\mu\,$.  Also remarkable is the fact that the mapping 
completely covers  both theories  in  the conformal (massless) limit of the
boundary field.

In order to consistently define a quantum field theory in $AdS$ space 
one actually needs a compactification of this space.
This way one is able to impose appropriate boundary conditions  
and avoid the loss or gain of information at spatial infinity in finite 
times and thus have a well defined Cauchy problem. 
This was established in \cite{QAdS1,QAdS2} in the context of 
global coordinates (these coordinates  have finite ranges).
  
Anti-de Sitter spacetime of $\,n+1$ dimensions can be represented\cite{Pe,Malda2} 
as the hyperboloid $X_0^2 + X_{n+1}^2 - \sum_{i=1}^n X_i^2\,=\,\Lambda^2 $
with $\,\Lambda\,=\,$ constant embedded in a flat $n+2$ dimensional space with metric 
$\, ds^2_{n+2}\,=\, - d X_0^2 - dX_{n+1}^2 + \sum_{i=1}^n dX_i^2\,$.
The so called Poincar\'e coordinates $\,z \,,\, \vec x\,,\,t\,$ are introduced by
\begin{eqnarray}
\label{Poincare}
X_0 &=& {1\over 2z}\,\Big( \,z^2\,+\,\Lambda^2\,
+\,{ \vec x}^2\,-\,t^2\,\Big)\,\,,
\nonumber\\
X_i &=& {\Lambda x^i \over z}\,\,\,\,\,,\,\,\,\,\,\,\,\,\,\,\,\,\,
X_{n+1} \,\,=\,\, {\Lambda t \over z}\,\,,
\nonumber\\
X_n &=& - {1\over 2z}\,
\Big( \,z^2\,-\,\Lambda^2\,+\,{\vec x}^2\,-\,t^2\,\Big)\,,
\end{eqnarray}

\noindent where $\vec  x \,=\, (x^1 , x^2 , ..., x^{n-1})\,$ with 
$ -\infty < x^i < \infty\,$ , $ -\infty < t < \infty\,$ and 
 $0 \le z < \infty $. In this case the $\,AdS_{n+1}\,$ measure with 
Lorentzian signature reads
\begin{equation}
\label{metric}
ds^2=\frac {\Lambda^2 }{ z^2}\Big( dz^2 \,+(d\vec x)^2\,
- dt^2 \,\Big)\,.
 \end{equation}
\noindent 

In recent articles \cite{BB1,BB2} we investigated the quantization of scalar 
fields  in the $AdS$ bulk in terms of Poincare coordinates, taking into account the
need of compactification of the space.
The $AdS$ boundary  corresponds to the region $\,z\,=\,0\,$ 
described by usual Minkowski coordinates $\vec x$ , $t\,$ plus 
a ``point'' at infinity ($z\,\rightarrow\,\infty\,$).
This point belongs to the boundary  in global coordinates
and must be added to the space in order to
find the appropriate compactification.
As discussed in \cite{BB1,BB2} this compactified $AdS$ space can not be completely 
represented in just one set of Poincar\'e coordinates. 
So one needs to introduce two coordinate charts in order to 
represent the compactified (in the axial $z$ direction) $AdS$ space. 
Each chart stops at some value of its $z$ coordinate.
The necessity of cutting this axial coordinate has 
the non trivial consequence that the field spectrum is discrete
in the $z$ direction as one should expect from a compact 
dimension. This reduces the dimensionality of the bulk space of states 
and makes it possible to find a one to one mapping  into the  boundary  states.
Note that one chart can be taken arbitrarily large in order to describe as much
of the $AdS$ space as wanted.

Let us  consider a massive scalar field $\Phi$ in the $\,AdS_{n+1}\,$
spacetime described by these coordinates with action
\begin{equation}
\label{action1}
I[\Phi ]\,=\, {1\over 2} \int d^{n+1}x \sqrt{g}\,
\left(\partial_\xi \Phi \, \partial^\xi \Phi
+m^2\,\Phi^2 \right)
\,\,,
\end{equation}
  
\noindent where we take $x^0\,\equiv\,z\,,\,x^{n+1}\,\equiv\,t\,$,
$\sqrt{g}\,=\,(x^0)^{-n-1}\,$ and $\xi\,=\,0,1,...,n+1\,$.

We consider a  Poincare chart in $AdS_{n+1}$ with\footnote{The $AdS_3$ case 
has some peculiarities and should be discussed separately.}  $n \ge 3$
given by 
$0\,\le\,z\,\le R\,$, where we will take $R$ to be 
arbitrarily large (but finite) in order to take as much of the $AdS$ space as we want.
The solutions of the classical equations of motion implied by the action
(\ref{action1}) can be used to construct quantum fields in this region 
giving\cite{BB1,BB2} 
\begin{equation}
\label{QF}
\Phi(z,\vec x,t)\,=\,\sum_{p=1}^\infty \,
\int { d \vec k \over (2\pi)^{n-1}}\,
{z^{n/2} \,J_\nu (u_p z ) \over R w_p(\vec k ) 
\,J_{\nu\,+\,1} (u_p R ) }
\lbrace { {\bf a}_p(\vec k )\ }
 e^{-iw_p(\vec k ) t +i\vec k \cdot \vec x}\,
\,+\,\,c.c.\rbrace\,,
\end{equation}

\noindent where $\vec k = (k_1 ,..., k_{n-1})\,$, \, 
$w_p(\vec k ) \,=\,\sqrt{ u_p^2\,+\,{\vec k}^2}$\, ,\, 
$u_p$ are such that $J_\nu(u_p R)=0$\, 
with $\nu=\frac 12\sqrt{n^2+m^2}$\,  \, 
and $c.c.$\, means complex conjugate. The operators 
${\bf a}_p\, ,\;{\bf a}^{\dagger}_p \,$ satisfy the commutation relations
\begin{equation}
\label{canonical1}
\Big[ {\bf a}_p(\vec k )\,,\,{\bf a}^\dagger_{p^\prime}({\vec k}^\prime  )
\Big]\,=\, 2\, (2\pi)^{n-1} w_p(\vec k )   
\delta_{p\,  p^\prime}\,\delta^{n-1} (\vec k -
{\vec k}^\prime )\,.
\end{equation}

On the $n$ dimensional boundary $ z = 0 $ we  consider quantum scalar fields
with a mass~$\mu$:
\begin{equation}
\label{BF}
\Theta_\mu ( \vec x ,t)\,=\,{1\over (2\pi)^{n-1}}
\int^{\infty}_{-\infty} { d \vec K  \over 2 w(\vec K ) }\,
\lbrace { {\bf b}( \vec K )\ }
 e^{-iw(\vec K ) t +i \vec K \cdot  \vec x }\,
\,+\,\,c.c.\rbrace\,,
\end{equation}

\noindent where $\vec K = (K_1, ..,K_{n-1})\,$ ,\, 
$ w(\vec K ) = \sqrt{ {\vec K}^2 + \mu^2}$\, and the creation-annihilation
operators satisfy the canonical algebra
\begin{equation}
\label{canonical2}
\Big[ {\bf b}( \vec K )\,,\,{\bf b}^\dagger ({ \vec K }^\prime  )
\Big]\,=\, 2 (2\pi)^{n-1} \,w( \vec K ) \delta ( \vec K -
{ \vec K}^\prime ) \,.
\end{equation}

\noindent Note that $\vec K$ and $\vec k$ have the same dimensionality
once we separate the component $u_p$ of the bulk momentum which is discrete.

In order to establish a correspondence between these two theories we use 
generalized spherical coordinate systems for representing 
both boundary and bulk momentum variables 
$\vec K \,=\,(K,\tilde\phi,{\tilde\theta}_\ell)$
and $\vec k \,=\, (k,\phi,\theta_\ell)$ respectively  
where $ K = \vert \vec K \vert \,$, $\, k = \vert \vec k \vert $
and $\ell = 1,..., n-3\,$. 
So we rewrite the phase space volume elements as 
\begin{eqnarray}
d\vec K &=& K^{n-2} d K \, d {\tilde \Omega}^{n-1} \nonumber\\
d\vec k &=& k^{n-2} d k \, d \Omega^{n-1}\,,
\end{eqnarray}

\noindent where $d {\tilde \Omega}^{n-1}\,$, $d \Omega^{n-1}\,$ 
are the infinitesimal elements of solid 
angle in $n-1$ dimensions for respectively  boundary and bulk.

Now, using this spherical coordinate representation, we introduce a 
sequence of energy scales $\epsilon_1 , \,\epsilon_2,\,...$
and split the operator $\Theta_\mu$ as 
\begin{eqnarray}
\label{BF2}
\Theta_\mu ( \vec x ,t) & = &
{1\over (2\pi)^{n-1}}
\int^{\epsilon_1}_{0} { K^{n-2}dK \over 2 w(K) }\int d {\tilde \Omega}^{n-1}
\lbrace  {\bf b}( \vec K )
 e^{-iw(K) t +i \vec K \cdot \vec x}
+  c.c.\rbrace\,
\nonumber\\
&+&{1\over (2\pi)^{n-1}}
\int^{\epsilon_2}_{\epsilon_1} {K^{n-2}dK \over 2 w(K) } 
\int d {\tilde \Omega}^{n-1}
\lbrace { {\bf b}( \vec K )\ }
 e^{-iw(K) t +i \vec K \cdot  \vec x}
+c.c.\rbrace \nonumber\\
&+& ...\,\,\,.
\end{eqnarray}

Then with a suitable mapping one can relate each of the $\Theta_\mu$
integrals above with the integral of the bulk field $\Phi$, eq.(\ref{QF}),
 over $d \vec k $ for a fixed $u_p$. Considering first the  interval 
$0 \le  K \le \epsilon_1 $  and $ p = 1 $ we introduce 
relations between the creation-annihilation operators of both theories.
We assume that  $ k$ is some  function of $ K$ and that the angular part of 
the mapping is trivial so that the same set of angular coordinates 
are used for bulk and boundary momenta. 
We choose
\begin{eqnarray}
K^{\,{n-2}\over 2} \,{\bf b}( K , \phi , \theta_\ell ) 
&=& k^{\,{n-2}\over 2}\,{\bf a}_1 (  k, \phi , \theta_\ell  ) \nonumber\\
K^{\,{n-2}\over 2}\,{\bf b}^\dagger ( K , \phi , \theta_\ell ) 
&=& k^{\,{n-2}\over 2}\,{\bf a}^\dagger_1 (  k, \phi , \theta_\ell    )\,,
\end{eqnarray}

\noindent where the moduli of the  momenta are mapped onto each other through
\begin{equation}
 k = g_1 (K,\mu)\,.
\end{equation} 

Requiring  that the canonical commutation relations 
(\ref{canonical1},\ref{canonical2}) are consistent with the above relations 
we find that the function $g_1$ is of the form:
\begin{equation}
g_1 (K,\mu) = {1\over 2} \,{u_1^2 C_1(\mu) 
			\over ( K + \sqrt{K^2 + \mu^2})}
-{1\over 2} { K + \sqrt{K^2 + \mu^2}\over C_1(\mu)}\,,
\end{equation}

\noindent where $C_1(\mu)$ is an arbitrary integration constant, for a given $\mu$. 
In order to have  $k \ge 0$ we put
\begin{equation}
C_1(\mu) =  {\epsilon_1  +\sqrt{\epsilon_1^2 + \mu^2 }\over u_1}
\end{equation}

\noindent so that the maximum value of $k=g_{1}(K,\mu)$ corresponds to 
$K = 0 $ and is given by
\begin{equation}
\lambda_1 =  \frac 12 u_1 
\left({\epsilon_1  +\sqrt{\epsilon_1^2 + \mu^2 }\over \mu}
-{\mu \over \epsilon_1  +\sqrt{\epsilon_1^2 + \mu^2 }}\right).
\end{equation}

Then, for the other intervals $\epsilon_{i-1} <  K \le \epsilon_i $ , 
that we put in correspondence with  $u_i$,
we introduce similarly the relations
\begin{eqnarray}
 {\bf b}( K , \phi , \theta_\ell ) &=& \Big[ { K^2 + \mu^2 \over 
(K - \epsilon_{i-1})^2 + \mu^2 }\Big]^{1\over 4}\,
\Big[ { g_i(K,\mu) \over K}\Big]^{\,{n-2}\over 2}
\,{\bf a}_i ( g_i (K,\mu) , \phi , \theta_\ell) 
\,\nonumber\\
{\bf b}^\dagger ( K , \phi , \theta_\ell ) &=&
\Big[ { K^2 + \mu^2 \over 
(K - \epsilon_{i-1} )^2 + \mu^2 }\,\Big]^{1\over 4}\,
\Big[ {g_i (K, \mu)\over K}\Big]^{\,{n-2}\over 2}\,
{\bf a}^\dagger_i ( g_{i} (K,\mu) , \phi , \theta_\ell  )  \,,
\end{eqnarray}

\noindent with $ k = g_i (K,\mu) $ and again we impose that the canonical relations 
(\ref{canonical1},\ref{canonical2}) are preserved, finding 
\begin{equation}
g_i (K,\mu) = {u_i \over 2} \,\left[ {\Delta\epsilon_i 
 +\sqrt{(\Delta\epsilon_i)^2 + \mu^2 } 
		\over K - \epsilon_{i-1} + 
\sqrt{(K - \epsilon_{i-1})^2 + \mu^2}}
- {K - \epsilon_{i-1} + \sqrt{(K - \epsilon_{i-1})^2 + \mu^2}\over 
\Delta\epsilon_i +\sqrt{(\Delta\epsilon_i)^2 + \mu^2 }}\right]\,,
\end{equation}

\noindent where $\Delta\epsilon_i=\epsilon_i-\epsilon_{i-1}$, 
so that $g_i ( \epsilon_i,\mu)=0$. 
The maximum  for $g_{i}(K,\mu)$ happens for $K=\epsilon_{i-1}$ 
and is given by
\begin{equation}
\label{lambdai}
\lambda_i =  \frac 12 u_i
\left(
{\Delta\epsilon_i+\sqrt{(\Delta\epsilon_i)^2+\mu^2}\over \mu}
-{\mu\over\Delta\epsilon_i+\sqrt{(\Delta\epsilon_i)^2+\mu^2}}
\right)\,.
\end{equation}

\noindent  
Note that the $\lambda_i$ are different in general depending on $u_i$
and $\Delta\epsilon_i$. The $u_i$ are related to the zeros of the Bessel
functions and obey the ordering $u_i > u_{i-1}$, but the intervals 
$\Delta\epsilon_i$ are of arbitrary size by construction.
This mapping between the momenta $K$ and $k$ is illustrated in. Fig.~1.

So we have established a correspondence between the states of scalar 
fields in $AdS$ bulk (massive or not) with massive scalar fields on 
its boundary. 

An important feature of this correspondence is that the boundary theory 
(which has an infrared cutoff $\,\mu\,$)
is mapped into a bulk theory with  ultraviolet
cutoffs $\lambda_i$ given by eq. (\ref{lambdai}) for each value of $\,u_i\,$.
Note that a  small $\mu  $  corresponds to  large $ \lambda_i $ (with a leading 
order term $\sim 1/\mu$). So  that we find explicitly a duality of the regimes UV-IR 
in the bulk/boundary mapping.


\
\setlength{\unitlength}{0.08in}
\vskip 3.5cm
{\begin{picture}(0,0)(13,0)
\rm
\put(33,-1){\vector(0,1){18}}
\put(34,17){$k$}
\put(32.5,12.5){\line(1,0){1}}
\put(29,12){$\lambda_2$}
\put(29,9.5){$\lambda_1$}
\put(32.5,15){\line(1,0){1}}
\put(29,14.5){$\lambda_3$}
\put(32,0){\vector(1,0){36}}
\put(45,-2){$\epsilon_1$}
\put(60,-2){$\epsilon_2$}
\put(68,-2){$K$}
\bezier{300}(33.1,9.9)(35,5)(46,0)
\put(33,10){\circle*{.6}}
\put(46,0){\circle*{.6}}
\put(36,7){$g_1(K,\mu)$}
\bezier{400}(46.1,12.4)(49,6)(61,0) 
\put(46,12.5){\circle{.6}}
\put(61,0){\circle*{.6}}
\put(50,8){$g_2(K,\mu)$}
\bezier{100}(61.1,14.9)(63,12)(66,11) 
\put(61,15){\circle{.6}}
\multiput(46,0)(0,2.2){6}{\line(0,1){1}}
\multiput(61,0)(0,2){8}{\line(0,1){1}}
\end{picture}
\vskip 1.5cm
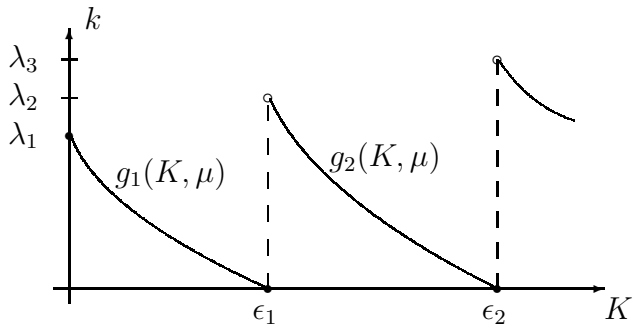
\noindent Figure 1: {\sl The mapping between the boundary 
momentum $K$ and bulk momentum $k$ for 
the case of a massive boundary theory. 
Finite intervals on $K$
are mapped into intervals for $k$ with cutoffs $\lambda_i$
for each $u_i$. }
\vskip 0.5cm

The   mapping of massive boundary fields into bulk scalar fields implies a direct 
relation between the corresponding quantum states

\begin{equation}
\vert  {\vec K}_i , \mu \rangle \,\,\,
\leftrightarrow \,\,\,\vert \vec k , u_i \rangle ,
\end{equation}

\noindent with $ \vec K_i = (K_i , \phi, \theta_\ell )$ being a momentum with modulus
$\epsilon_{i-1} <  K_i \le \epsilon_i \,$ and 
$ \vec k = (k , \phi, \theta_\ell )$  with modulus 
$0 \le k < \lambda_i $.
This is a realization of the Holographic principle in terms of quantum states
and it exhibits the Ultraviolet-Infrared duality expected from  the bulk 
boundary correspondence\cite{HOL3}.

Now we focus  on the important limiting case of a massless (conformal) boundary theory.
First we note that the first interval will be taken as $0 < K \le \epsilon_1 $,
excluding the state of $K = 0$ that is not physically relevant in this massless case.
The other intervals are taken again as $\epsilon_{i-1} <  K \le \epsilon_i $ 
as in the massive case. Here we find ( $\epsilon_0 \equiv 0$)

\begin{equation}
g_i (K) = {u_i \over 2} \,\Big[  { \Delta\epsilon_{i} \over K - \epsilon_{i-1}}
-  {K - \epsilon_{i-1}\over \Delta\epsilon_{i}}\Big]\,,
\end{equation}

\noindent  such that $g_i (\epsilon_i) = 0$.
Note that the maximum  for $g_{i}(K)$ also happens for $K=\epsilon_{i-1}$ 
and is given by

\begin{equation}
\lambda_i =  \frac 12 u_i
\left.\left(
{\Delta\epsilon_i\over \mu}-{\mu\over\Delta\epsilon_i}
\right)\right|_{\mu\to 0} \longrightarrow \infty\,.
\end{equation}
 
So we find out that when the boundary theory
is conformal the whole phase space of the bulk  (without any UV cutoff)
is mapped in the whole phase space of the boundary  (with the exception of the 
state of zero momentum that has no Physical content). 
This one to one mapping between the momenta $K$ and $k$ 
is represented in Fig.~2. 


\
\setlength{\unitlength}{0.08in}
\vskip 3.5cm
{\begin{picture}(0,0)(13,0)
\rm
\put(33,-1){\vector(0,1){18}}
\put(34,17){$k$}
\put(32,0){\vector(1,0){36}}
\put(45,-2){$\epsilon_1$}
\put(60,-2){$\epsilon_2$}
\put(68,-2){$K$}
\bezier{400}(33.3,15)(34,1)(46,0)
\put(46,0){\circle*{.6}}
\put(37.5,5){$g_1(K)$}
\bezier{400}(46.3,15)(46.5,1)(61,0) 
\put(61,0){\circle*{.6}}
\put(51,5){$g_2(K)$}
\bezier{200}(61.3,15)(61.4,8)(65,5) 
\multiput(46,0)(0,2){8}{\line(0,1){1}}
\multiput(61,0)(0,2){8}{\line(0,1){1}}
\end{picture}
\vskip 1.5cm
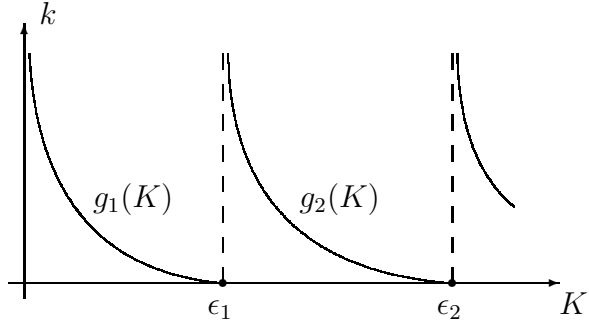
\noindent Figure 2: {\sl The mapping between the boundary 
momentum $K$ and bulk momentum $k$ for 
the case of a conformal boundary theory. 
Every finite interval on $K$
is mapped into an infinite interval for $k$ 
(corresponding to each value of $u_i$),
so that the bulk phase space is completely covered by this mapping.}
\vskip 0.5cm

In the conformal case we found  a direct mapping of boundary/bulk  quantum states:

\begin{equation}
\vert {\vec K}_i  \rangle \,\,\leftrightarrow \,\,\vert \vec k  , u_i  \rangle\,,
\end{equation}

\noindent where again $\epsilon_{i-1} <  K_i \le \epsilon_i \,$ but now 
$0 \le k < \infty $ without any ultraviolet cutoff.

The correlation functions for the conformal boundary theory can be calculated 
\cite{FGG} directly from the boundary fields, eq. (\ref{BF}) 

\begin{equation}
\langle \Theta_{0}
( x )\Theta_{0}
( x^\prime )\rangle\sim { 1 \over ( x \,-\, x^\prime \,)^{2d} }\,,
\end{equation}

\noindent where $x = (\vec x , t)$ , $x^\prime = (\vec x^\prime  , t^\prime )$
and $d = (n -2)/2 $ is the conformal dimension for the scalar field 
$\Theta_0 \equiv \Theta_{\mu = 0}$ defined in the boundary of  $AdS_{n+1}$.

Let us now comment on the differences between our approach and
that of the \break AdS$_{n+1}$/CFT$_{n}$ correspondence \cite{Malda,GKP,Wi}.
In that case bulk scalars of mass $m$ are mapped into boundary composite 
operators of conformal dimension $( n + \sqrt{n^2 + 4 m^2})/2$. 
It is interesting to note that $m^2$ can be negative subjected to a lower 
bound $m^2\ge -n^2/4$ \cite{QAdS2,Wi}, so that the conformal dimension is 
$\ge n/2$.
This dimension will not match that of our boundary field because we
considered a simpler situation of bulk and boundary scalar theories.
However with this simple model we found a direct one to one mapping 
between quantum states. 
   
We expect that our mapping could  be generalized to other  
fields if one starts with  appropriate expansion for the boundary operators.
This would enlarge the mechanism proposed here possibly
allowing the inclusion of composite operators. 
In that case the relation between such a mapping and the 
AdS/CFT correspondence would be closer.

Finally we point out that once established a  one to one mapping 
between  bulk and boundary quantum states it is possible to relate their 
entropies in the same way. So the entropy area law would 
be a  consequence of this mapping, at least for the system of 
scalar fields analyzed here.

\section*{Acknowledgments} 
We would to thank Marcelo Alves for interesting discussions. 
The authors are partially supported by CNPq, FINEP and FAPERJ 
- Brazilian research agencies.



\end{document}